# Distributed UAV-enabled zero-forcing cooperative jamming scheme for safeguarding future wireless networks

X. A. Flores Cabezas; D. P. Moya Osorio; and M. Latva-aho
Centre for Wireless Communications, University of Oulu, Finland
Email: [xavier.florescabezas, diana.moyaosorio, matti.latva-aho]@oulu.fi

*Abstract*—In this work, we investigate the impact of two cooperative unmanned aerial vehicle (UAV)-based jammers on the secrecy performance of a ground wireless network in the presence of an eavesdropper. For that purpose, we investigate the secrecy-area related metrics, Jamming Coverage and Jamming Efficiency. Moreover, we propose a hybrid metric, the so-called Weighted Secrecy Coverage (WSC) and a virtual distributed multiple-input-multiple-output (MIMO)-based zero-forcing precoding scheme to avoid the jamming effects on the legitimate receiver. For evaluating these metrics, we derive a closed-form position-based metric, the secrecy improvement. Our mathematical derivations and comparative simulations show that the proposed zero-forcing scheme leads to an improvement on the secrecy performance in terms of the WSC, and provides conditions for improvement of Jamming Efficiency. They also show positioning trends on the UAVs over a fixed orbit around the legitimate transmitter as well as power allocation trends for optimal secrecy.

*Keywords—cooperative jamming, physical layer security, UAV communications, zero-forcing precoding.*

## I. INTRODUCTION

The envisioned fully automated and intelligent networks will be only concretised in the sixth generation (6G) of wireless networks. 6G will further develop 5G systems by introducing new disruptive technologies jointly with a pervasive use of artificial intelligence to provide highly intelligent, efficient and secure services [1]. To fulfil the ambitious goals raised for 6G, aerial platforms such as unmanned aerial vehicles (UAV) are expected to be an integral part of the 6G architecture. Due to their mobility, ease of deployment and the high probability of establishing line-of-sight (LoS) links, UAV-enabled communications are attractive for different use cases such as disaster prevention and recovering, enhancement of coverage for rural areas and hotspots, offloading terrestrial base stations (BSs) in dense urban areas, among others [2].

The flexibility and dynamism offered by the use of UAVs bring both opportunities and challenges for these networks. Particularly, the predominance of LoS channels makes the network prone to eavesdropping attacks. However, security can be improved by exploiting the controllable mobility of UAVs. Thus, information security is one of the most prominent issues to be tackled in UAV-enabled networks [3]. In this sense, physical layer security (PLS) techniques have shown potential as a compelling option for traditional upper-layer cryptographic mechanisms [3]–[7]. For instance, in [4], the outage probability (OP) of a legitimate receiver and the intercept probability (IP) of an eavesdropper are evaluated by considering a UAV-based cooperative jammer for enhancing secrecy. Therein, a metric so-called intercept probability security region is proposed and maximised by optimising the 3D deployment and power of the UAV jammer, given an OP constraint. In [5], the secrecy outage probability and the average secrecy rate are analyzed for a UAV-based system, where a ground IoT device transmits confidential information to a UAV, while a random number of ground randomly located eavesdroppers try to leak information. The analysis is carried out for the case without and with a UAV that generates jamming signals to distract the eavesdroppers. Moreover, in [6], a UAV is used as a cooperative jammer for providing confidentiality to an air-to-ground (A2G) communication by considering partial information on the location of the eavesdropper. Also, UAV-enabled friendly jamming has been a topic of interest in industrial IoT security as in [8], where UAVs are used as cooperative jammers to disrupt eavesdroppers by considering specific eavesdropper appearance regions, where the positioning of various UAVs with directional antennas is analyzed to perform beamforming over the intended areas.

All in all, the PLS technique of sending artificial noise as friendly jamming has shown to greatly improve the secrecy performance of wireless networks. Therefore, inspired by the secrecy-related area performance metrics proposed in [9], herein we evaluate the impact of cooperative jamming, performed by UAV-enabled jammers, over the secrecy performance of the communication between a pair of legitimate nodes in the presence of an eavesdropper. Particularly, we evaluate the area-based metrics: Jamming Coverage and Jamming Efficiency, and we introduce the Weighted Secrecy Coverage (WSC) metric as a helpful tool to evaluate the impact of cooperative jamming on the enhancement of secrecy in UAV-enabled networks.

In this work, we propose a virtual distributed multiple-input-multiple-output (MIMO) transmission scheme with zero-forcing precoding in order to eliminate the impact of the jamming signal on the legitimate receiver by considering no knowledge about the position of the eavesdropper. This is done by considering two jamming UAVs and the legitimate transmitter Alice as geographically dislocated antennas part of a virtual transmitter and the legitimate receiver Bob and the eavesdropper Eve as geographically dislocated antennas part of a virtual receiver, composing the virtual MIMO system, where a precoding matrix is applied to the virtual transmitter that considers reliable channel state information (CSI) on the channel between the virtual transmitter and Bob in order to suppress the jamming signals at this node. Then, we analyze the secrecy performance of the system making use of the previously mentioned area secrecy metrics that do not consider information about Eve's channel or location and provide a comparison to the case where no precoding is used. To this end, we also derive the jamming improvement metric, which is an analogous variation of the one proposed in [9], in closed form.

## II. SYSTEM MODEL

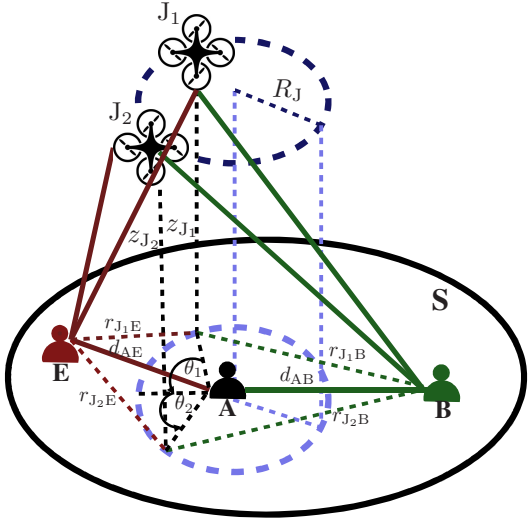

Fig. 1. System model.

Consider the system depicted in Fig. 1 composed by a legitimate ground transmitter, Alice (A) at position $(x_A, y_A)$; a legitimate ground receiver, Bob (B) at position $(x_B, y_B)$; a ground eavesdropper, Eve (E) at position $(x_E, y_E)$; and two UAV-enabled friendly jammers, $J_i$ with $i \in \{1, 2\}$ at positions $(x_{J_i}, y_{J_i})$. The legitimate ground nodes are single-antenna low complexity receivers which may be seen as IoT nodes engaging in machine-type communications. In this scenario, we consider that E is located at an arbitrary position of a circular space, $S$, and tries to leak confidential information from the communication between A and B. The ground links are assumed to undergo Rayleigh fading and are subject to additive white Gaussian noise (AWGN), with mean power $\sigma^2$. Therefore, the corresponding channel coefficients, $h_{AU}$ with $U \in \{B, E\}$, are independent complex circularly-symmetric Gaussian random variables with variance $\mathbb{E}\left[|h_{AU}|^2\right] = d_{AU}^{-\alpha}$, where $d_{AU}$ being the distance among a pair of nodes and $\alpha$ being the pathloss exponent for the ground links. Thus, the channel gains, $g_{AU} = |h_{AU}|^2$, follow an exponential distribution with parameter $\Omega_{AU} = d_{AU}^{\alpha}$. To enhance the secrecy performance of the system, the two friendly jammers are positioned so that they introduce artificial pseudo-noise in order to interfere E. Herein, we assume that the aerial jammers move in a three-dimensional space while their projection into the ground plane is confined to a circumference of radius $R_J$ around A, so we express their positions in cylindrical coordinates $(r_{J_i}, \theta_{J_i}, z_{J_i})$. The A2G links, between the jammers and B or E, can experience either line-of-sight (LoS) or non-line-of-sight (NLoS) propagation. Therefore, the average pathloss for the jamming links are given by [10]

$$L_{J_iU} = \left(z_{J_i}^2 + r_{J_iU}^2\right)^{\frac{\alpha_J}{2}} (P_{\text{LoS}}\eta_{\text{LoS}} + P_{\text{NLoS}}\eta_{\text{NLoS}}),$$

where $z_{J_i}$ is the height of $J_i$, $r_{J_iU}$ is the distance from node U and the projection on the plane of $J_i$, $\alpha_J$ is the pathloss exponent for the A2G links, $\eta_{\text{LoS}}$ and $\eta_{\text{NLoS}}$ are the attenuation factors for the LoS and the NLoS links, respectively. Also, $P_{\text{LoS}}$ and $P_{\text{NLoS}}$ are the probabilities of LoS and NLoS connection, respectively given by

$$P_{\text{LoS}} = \frac{1}{1 + \psi \exp\left(-\omega\left[\frac{180}{\pi}\tan^{-1}\left(\frac{z_{J_i}}{r_{J_iU}}\right) - \psi\right]\right)}, \quad (1)$$

$$P_{\text{NLoS}} = 1 - P_{\text{LoS}}, \quad (2)$$

where $\psi$ and $\omega$ are environmental constants [11], [12]. Thus we express the deterministic A2G channel coefficients as $h_{J_iU} = (\sqrt{L_{J_iU}})^{-1}$, and the channel gains as $g_{J_iU} = |h_{J_iU}|^2$ with $i \in \{1, 2\}$. Herein, we will evaluate the impact of the friendly jamming signal over the secrecy performance of the proposed system in terms of two secrecy metrics proposed in [9], the Jamming Coverage and the Jamming Efficiency, which will be described next. In that case, we are interested in analysing the gain obtained by the friendly jammer. Therefore, the received signal-to-noise ratios (SNRs) at B and E, are given by $\gamma_B = a_j \cdot g_{AB}, \gamma_E = b_j \cdot g_{AE}$, with $j \in \{\text{NJ}, \text{J}\}$, for the scenarios with and without friendly jamming, respectively. Thus, $a_{\text{NJ}} = b_{\text{NJ}} = \gamma_A$, with $\gamma_A$ and $\gamma_{J_i}$ being the transmit SNRs at A and $J_i$, with $\gamma_A = \frac{P_A}{\sigma^2}$ and $\gamma_{J_i} = \frac{P_{J_i}}{\sigma^2}$, and $P_A$ and $P_{J_i}$ are the transmit powers at Alice and the jamming power at $J_i$, respectively. Accordingly, for the friendly jamming case, $a_J$ and $b_J$ are given by

$$a_J = \frac{\gamma_A}{1 + g_{J_1B}\gamma_{J_1} + g_{J_2B}\gamma_{J_2}}, \quad (3)$$

$$b_J = \frac{\gamma_A}{1 + g_{J_1E}\gamma_{J_1} + g_{J_2E}\gamma_{J_2}}. \quad (4)$$

### A. Zero-forcing precoding scheme

We assume the proposed system as a distributed virtual MIMO system with three-element input $s[m]$ and two-element output $y[m]$ for the $m$-th symbol over a channel matrix $H$. The inputs are given by $s[m] = [s_A[m]\ s_{J1}[m]\ s_{J2}[m]]^T$, where $s_A[m]$ is the $m$-th symbol sent by Alice and $s_{J1}[m]$ and $s_{J2}[m]$ are the $m$-th generated pseudo-random artificial noise sent by $J_1$ and $J_2$, respectively, all of which follow a standard normal distribution. The outputs are given by $y[m] = [y_B[m]\ y_E[m]]^T$ where the elements are the received signals at Bob and Eve, respectively. Then, the channel matrix $H$ is expressed as

$$H = \begin{bmatrix} h_{AB} & h_{J_1B} & h_{J_2B} \\ h_{AE} & h_{J_1E} & h_{J_2E} \end{bmatrix}, \quad (5)$$

The input signal $s[m]$ is sent through a power assigning matrix $P = diag\{\sqrt{P_A}, \sqrt{P_{J_1}}, \sqrt{P_{J_2}}\}$ and then through a zero-forcing precoding matrix $Q$. Then the received signals are given by:

$$y[m] = Hx[m] + w[m] = HQPs[m] + w[m], \quad (6)$$

where $w[m] = [w_B[m]\ w_E[m]]^T$ is a noise vector of i.i.d. elements that follow a Gaussian distribution with variance $\sigma^2$. We also assume no knowledge on Eve's channel, then the precoding matrix is designed to eliminate the jamming effect at B. In that case, we assume that the CSI of the links between the jammers and B is perfectly known by J1 and J2, as there is a reliable channel between them, then full coordination is possible and both jammers send the same pseudo-random sequence $s_{J1}[m] = s_{J2}[m] = s_J[m]$ with the same power

$P_{J_1} = P_{J_2} = P_J$. This can be achieved while avoiding leakage to Eve by having a separate set of directional antennas between the UAVs or by more elaborate schemes involving cooperation from Alice and Bob as shown in [13]. Then, $Q$ is given by $Q = diag\{1, h_{J_2B}, -h_{J_1B}\}$ and $y[m]$ is expressed as

$$y[m] = \begin{bmatrix} h_{AB}\sqrt{P_A}s_A[m] + w_B[m] \\ h_{AE}\sqrt{P_A}s_A[m] + h_{INT}\sqrt{P_J}s_J[m] + w_E[m] \end{bmatrix}, \quad (7)$$

with $h_{INT} = h_{J_1E}h_{J_2B} - h_{J_2E}h_{J_1B}$ being the effective deterministic channel response of the joint cooperative jamming from J1 and J2 with corresponding channel gain $g_{INT} = |h_{INT}|^2$. By applying the precoding scheme, the impact of the jamming signal at B has been removed. Then, we have that

$$a_J = \gamma_A, \quad (8)$$
$$b_J = \frac{\gamma_A}{1 + g_{INT}\gamma_J}. \quad (9)$$

### III. Performance Analysis

In this Section, we focus our analysis on secrecy metrics that allow us to evaluate the impact of friendly jamming on the secrecy performance of the proposed system, while considering the fact that the location of E is unknown, but within the working area S. We resort to the secrecy metrics proposed in [9], Jamming Coverage and Jamming Efficiency, and we propose a hybrid metric, so-called Weighted Secrecy Coverage.

These metrics are expressed in terms of the well-known secrecy outage probability (SOP), which is the probability that the secrecy capacity falls below a certain secrecy rate $R_S$ [14], given by

$$\text{SOP} = \Pr[C_S < R_S], \quad (10)$$

where $C_S$ is the secrecy capacity defined as the maximum achievable rate to maintain a confidential communication [14], that is, the maximum value between 0 and the difference between the capacities of the legitimate channel $C_B$ and the wiretap channel $C_E$ given by [14]

$$C_S = [C_B - C_E]^+ = \left[\log_2\left(\frac{1+\gamma_B}{1+\gamma_E}\right)\right]^+, \quad (11)$$

with $[X]^+ = \max[X, 0]$. Then, we define $\overline{\Delta}$ as the secrecy improvement due to friendly jamming given as the ratio of the attained probability of achieving secrecy considering friendly jamming and the attained probability of achieving secrecy without friendly jamming, which is an analogous metric to the one presented in [9]

$$\overline{\Delta} = \frac{1 - \text{SOP}_J}{1 - \text{SOP}_{NJ}}. \quad (12)$$

From (12), we can notice that $\overline{\Delta}$ provides information on the impact of jamming at a certain position for Eve. In this reason, for the positions where $\overline{\Delta} < 1$, the presence of jamming worsens the secrecy performance; whereas, if $\overline{\Delta} > 1$, the presence of jamming improves the secrecy performance. Analysing this for every point in $S$, the Jamming Coverage can be defined as the total area that has a value of $\overline{\Delta} > 1$ [9], so that the presence of the jammers would cause a benefit on the secrecy of the system. Thus, the Jamming Coverage can be expressed as

$$\text{Coverage} = \iint_{\overline{\Delta}>1} dS. \quad (13)$$

On the other hand, the Jamming Efficiency is defined as the mean improvement in secrecy over the whole area $S$ [9] as

$$\text{Efficiency} = \frac{1}{|S|}\iint_S \overline{\Delta}\,dS. \quad (14)$$

This metric gives a measurement of the average impact of friendly jamming over the secrecy performance. Considering these two metrics and under the observation (after several simulations) that the jamming Efficiency value usually oscillates around 1, and does not fluctuate much, we can consider the jamming Efficiency as a weight factor for the jamming Coverage, thus we can define a hybrid metric, the Weighted Secrecy Coverage (WSC), as follows

$$\text{WSC} = \left(\iint_{\overline{\Delta}>1} dS\right)\left(\frac{1}{|S|}\iint_S \overline{\Delta}\,dS\right). \quad (15)$$

This metric considers the area where the secrecy is improved with the presence of friendly jamming joint with the mean improvement over the whole area. The WSC metric is completely determined by $\overline{\Delta}$, which is obtained for the proposed system as in the following Proposition.

**Proposition 1.** *The jamming improvement, $\overline{\Delta}$, for the proposed system model, is given by*

$$\overline{\Delta} = e^{\Omega_{AB}(2^{R_S}-1)\left(\frac{1}{a_J}-\frac{1}{a_{NJ}}\right)} \left(\frac{2^{R_S}\left(\frac{\Omega_{AB}}{\Omega_{AE}}\right)+1}{2^{R_S}\left(\frac{\Omega_{AB}}{\Omega_{AE}}\right)\left(\frac{b_J}{a_J}\right)+1}\right). \quad (16)$$

*Proof:* The proof is provided in Appendix A. ∎

#### A. Impact of zero-forcing precoding on secrecy metrics

Considering the parameters for the precoding scheme as shown in (8) and (9), $\overline{\Delta}$ is reduced to:

$$\overline{\Delta} = \frac{2^{R_S}\left(\frac{\Omega_{AB}}{\Omega_{AE}}\right)+1}{2^{R_S}\left(\frac{\Omega_{AB}}{\Omega_{AE}}\right)\left(\frac{1}{1+g_{INT}\gamma_J}\right)+1}, \quad (17)$$

which does not depend on the power allocated to Alice. Analysing the region where $\overline{\Delta} > 1$, for which jamming improves secrecy, basic algebraic derivation leads to the condition $g_{INT} > 0$, which is always true. Furthermore, the condition for $g_{INT} = 0$ is found to be $h_{J_1E}h_{J_2B} = h_{J_2E}h_{J_1B}$, and it reduces to $h_{J_1E} = h_{J_2E}$ when jammers are located symmetrical to Bob. Either way, this condition draws a 1-dimensional curve for the position of Eve in area $S$, which does not contribute to an area calculation. Thus, under the proposed precoding scheme, the secrecy is (practically) always improved regardless of the position of Eve. Then, the WSC is expressed as

$$\text{WSC} = \iint_S \overline{\Delta}\,dS = |S| \cdot \text{Efficiency}. \quad (18)$$

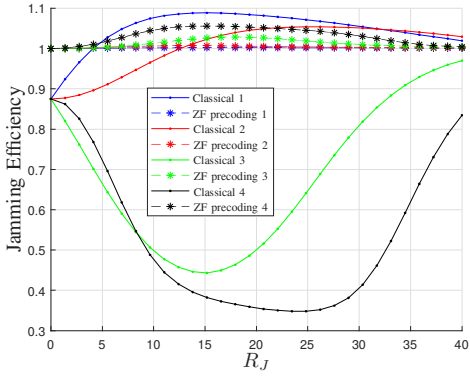

Fig. 2. Jamming efficiency at set angle values ($\theta_{J1}$, $\theta_{J2}$) (1: (45°, -45°), 2: (90°, -90°), 3: (135°, -135°), 4: (0°, 180°)) for varying values of surveillance radius $R_J$ of jammers around Alice, with and without considering the precoding scheme ($z_J = 13$, $d_{AB} = 20$, $\gamma_A$=10, $\gamma_{J1}$=5, $\gamma_{J2}$=5).

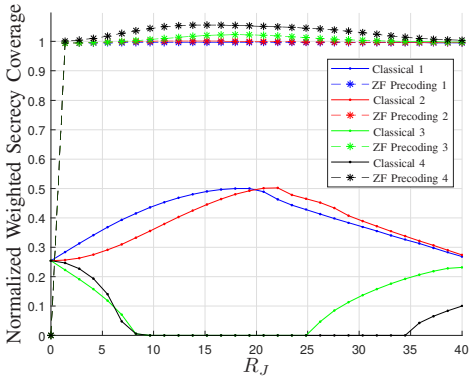

Fig. 3. Area-normalized Weighted Secrecy Coverage at set angle values ($\theta_{J1}$, $\theta_{J2}$) (1: (45°, -45°), 2: (90°, -90°), 3: (135°, -135°), 4: (0°, 180°)) for varying values of surveillance radius $R_J$ of jammers around Alice, with and without considering the precoding scheme ($z_J = 13$, $d_{AB} = 20$, $\gamma_A$=10, $\gamma_{J1}$=$\gamma_{J2}$=5).

## IV. NUMERICAL RESULTS AND DISCUSSION

In this Section, we evaluate the impact of the zero-forcing precoding cooperative jamming scheme over different system parameters on the secrecy performance of the proposed system. For the environmental and link constants, we assume a urban scenario with the following values: $\psi = 9.61$, $\omega = 0.16$, $\alpha = 0.3$, $\alpha_j = 0.3$, $\eta_{\text{NLoS}} = 20$ and $\eta_{\text{LoS}} = 1$ [11], [12]. The position of A is set at coordinates $(0, 0)$ and B is located at coordinates $(d_{AB}, 0)$, so the cylindrical coordinates of both jammers become $(R_J, \theta_{J1}, z_{J1})$ and $(R_J, \theta_{J2}, z_{J2})$, respectively with the reference starting from the left of A, $\theta_{J1}$ opening clockwise and $\theta_{J2}$ counterclockwise as is illustrated in Fig. 1. Also the jammers are considered to be located at the same fixed height $z_J$. For practicality purposes, we refer to the case where non-precoding scheme is applied as classical scheme.

Fig. 2 and Fig. 3 illustrate the Jamming Efficiency and the area-normalized WSC by assuming different values of surveillance radius around A $R_J$ for both schemes. For each radius, four angle configurations are tested. Configuration 1 through 3 locate both jammers symmetrically regarding the x axis, with angles of 45°, 90° and 135° while configuration 4 considers that one jammer is located directly behind Alice at 0° and the other is located directly in front at 180°. For cases with precoding scheme, note that the efficiencies never go below one unlike the classical scheme. For configurations 3 and 4, the efficiencies are always below one. However, it is observed that configuration 1 of the classical scheme attains better efficiency than the precoding scheme for all configurations after a certain value of $R_J$. Regarding the WSC, it can be observed that the precoding scheme has a better performance than the classical scheme regardless of the configuration or $R_J$ used. This is consistent with the mathematical results since the precoding scheme Coverage equals the whole area $S$, which is the dominant term in the WSC computation. Moreover, for the classical scheme, it is observed that there are values of $R_J$ for configurations 3 and 4 where the WSC reaches a value of zero, as the Jamming Coverage equals 0. From the results, it can be also observed that the optimal tendency of the angle allocation of the jammers using the precoding scheme in terms of the Jamming Efficiency is on configuration 4, that is directly behind and directly in front of Alice, which is different than the optimal tendency of the classical scheme which is under configuration 1, that is jammers behind Alice at 45° symmetrical to the horizontal axis.

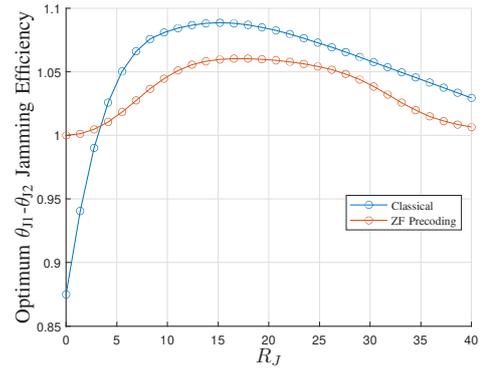

Fig. 4. Jamming efficiency at optimum angle values $\theta_{J1}$ and $\theta_{J2}$ for varying values of surveillance radius $R_J$ of jammers around Alice, with and without considering the precoding scheme ($z_J = 13$, $d_{AB} = 20$, $\gamma_A$=10, $\gamma_{J1}$=$\gamma_{J2}$=5).

For the following figures, a range of angles around A from -180° to 180° for both jammers are tested for each radius, and the ones that give the maximum secrecy values, which are shown in Fig. 5, are considered. It is clear that the WSC will always be higher in the proposed precoding scheme, so it is of interest to consider the optimal Efficiencies and angular positioning regarding WSC.

Fig. 4 illustrates the maximum Jamming Efficiency achieved for each value of $R_J$. Note that for values higher than $\approx 4$, the precoding scheme approach presents a worse Jamming Efficiency than the classical approach. This behaviour is due to the $g_{\text{INT}}$ term in (9), which is weighted by the channels between the jammers and Bob, for which the gains become smaller as the distance between them increase, which is the case for one of the jammers at higher radius length, as Fig. 5 suggests.

Fig. 5 presents the angles for optimal WSC. Note that, for the classical scheme, the angular distribution suggests that below a certain $R_J$ value (around 5), the optimal positioning for both jammers is located directly behind A. Above this, the optimal positioning for the jammers is to be located behind Alice, positioned symmetrically with respect to the x axis at different angles. For the precoding scheme, the optimal positioning for the jammers is when one jammer is consistently located directly in front of Alice and below a certain $R_J$ (around 9) the other jammer is located directly behind Alice, both of them forming a 180° angle. As $R_J$ approaches the middle point between Alice and Bob ($R_J$=10), the jammer located behind Alice starts to shift its positioning, moving from the back of Alice towards the front in increasing angles, while the jammer in front of Alice remains fixed at 180°.

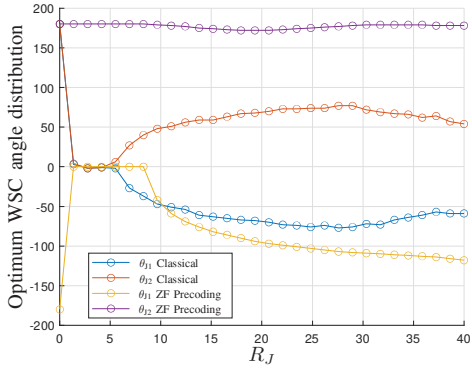

Fig. 5. Optimal WSC angle distribution of jammers $\theta_{J1}$ and $\theta_{J2}$ for varying values of surveillance radius $R_J$ around Alice, with and without considering the precoding scheme ($z_J = 13$, $d_{AB} = 20$, $\gamma_A=10$, $\gamma_{J1}=5$, $\gamma_{J2}=5$).

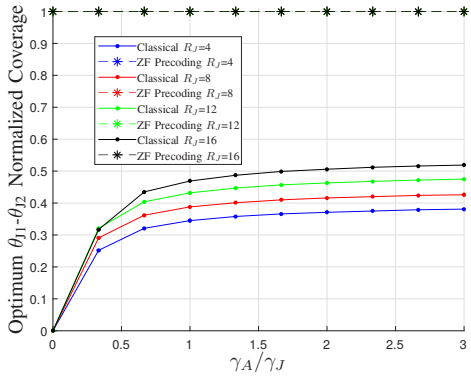

Fig. 6. Normalized coverage at optimum angle values $\theta_{J1}$ and $\theta_{J2}$ for varying values of power distribution $\gamma_A/\gamma_J$ between A and the jammers, with and without considering the precoding scheme ($z_J = 13$, $d_{AB} = 20$, $\gamma_T=20$).

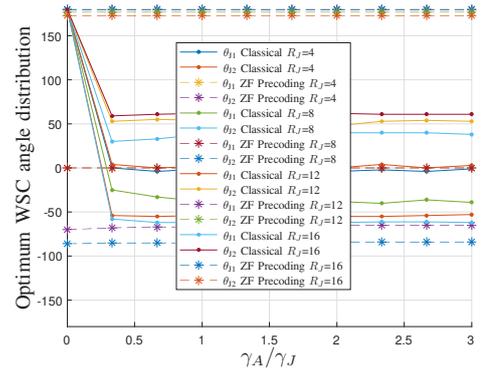

Fig. 7. Optimal WSC angle distribution of jammers $\theta_{J1}$ and $\theta_{J2}$ for varying values of power distribution $\gamma_A/\gamma_J$ between Alice and the jammers, with and without considering the precoding scheme ($z_J = 13$, $d_{AB} = 20$, $\gamma_T=20$).

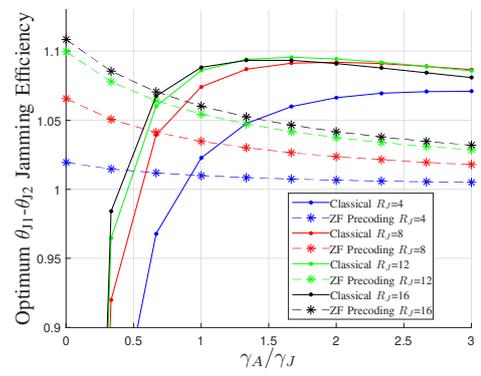

Fig. 8. Jamming efficiency at optimum angle values $\theta_{J1}$ and $\theta_{J2}$ for varying values of power distribution $\gamma_A/\gamma_J$ between Alice and the jammers, with and without considering the precoding scheme ($z_J = 13$, $d_{AB} = 20$, $\gamma_T=20$).

In the following figures, we evaluate the impact of the total power allocation on the secrecy metrics. We consider that a given total transmit power $P_T$, with $\gamma_T = P_T/\sigma^2$, is available to be divided between Alice and the jammers, and $\gamma_J = \gamma_{J1} + \gamma_{J2}$ with $\gamma_{J1} = \gamma_{J2}$. We determine the optimal angles to maximize the studied secrecy metrics for a range of power allocation ratios $\gamma_A/\gamma_J$, where the zero point represents all the power allocated to the jammers, and as it gets further to the right the allocation shifts in favor to Alice.

Figure 6 shows the results for the Jamming Coverage metric. Note that, for the precoding scheme, the Jamming Coverage remains the same regardless the power allocation, which is to be expected. In the classical scheme, for larger values of $R_J$ up to 16, the Jamming Coverage increases, and, for all the values of $R_J$, the same tendency in terms of power allocation is observed, where the more power is allocated to A, the better the Jamming Coverage.

In Fig. 7, the optimal angles, in terms of WSC are shown for different $R_J$ values, over a range of power allocations. Note that for a given $R_J$, the power allocation does not introduce significant fluctuations in the optimal angle distribution of the jammers, thus the surveillance radius presents more impact over the angle distribution of the UAVs rather than the power allocation.

Fig. 8 shows the Jamming Efficiency metric. Note that there are power allocation ratios for which the precoding scheme outperforms the classical scheme in terms of Jamming Efficiency. For every $R_J$ value, the precoding scheme performs better than the classical scheme for low values of allocation ratio, that is when more power is allocated to the jammers. However it seems that the power allocation point 0.5, that is equal allocation between the three transmitters (A and both jammers), is the highest allocation for which the precoding scheme performs equal or better than the classical scheme. Also, note that the maximum Efficiency value for the precoding scheme is achieved when no power is allocated towards Alice, which is not a case of interest. However we can see from this that the Efficiency curve is decreasing, which is to say it is higher the more power is allocated to the jammers and not to Alice.

### A. Impact of imperfect CSI

If imperfect CSI is assumed at the UAVs, the channel estimates between the jammers and Bob can be expressed as $\hat{h}_{J_iB} = h_{J_iB} + \delta_i$ with $i \in \{1,2\}$, where $\delta_i$ represent the estimation error term, which could be expressed as $\delta_i = p_i h_{J_iB}$ where $p_i$ is the proportional error regarding the true value of the channel response. Then, the imperfect CSI precoding matrix is given by $Q = diag\{1, \hat{h}_{J_2B}, -\hat{h}_{J_1B}\}$, and the received signal can be expressed as

$$y[m] = \begin{bmatrix} h_{AB}\sqrt{P_A}s_A[m] + \hat{h}_{INTB}\sqrt{P_J}s_J[m] + w_B[m] \\ h_{AE}\sqrt{P_A}s_A[m] + \hat{h}_{INTE}\sqrt{P_J}s_J[m] + w_E[m] \end{bmatrix}, \quad (19)$$

where $\hat{h}_{INTB} = h_{J_1B}h_{J_2B}(p_2 - p_1)$ and $\hat{h}_{INTE} = h_{INT} + p_2 h_{J_1E}h_{J_2B} - p_1 h_{J_2E}h_{J_1B}$. Note that if the proportional errors on both channels are the same, the impact of the jamming is removed at Bob. However, it persists at Eve. The analysis of this scenario is out of the scope of this work, and may be of future research interest.

## V. Conclusions

This work studied the impact of the proposed zero-forcing precoding scheme where the interference is nullified at Bob, making use of area-based secrecy metrics that consider no information on the position of Eve within a given area. The precoding scheme in general achieves higher secrecy performance than the classical scheme in terms of WSC, so it allows an increase in secrecy. In terms of Coverage the precoding scheme always performs better than the classical scheme, while in terms of Efficiency it performs the same or better when the available power is distributed evenly between Alice and the two UAVs, or as more power is allocated towards the UAVs. The best UAV allocation for this precoding scheme is simpler than the classical scheme and requires updating the position of only one of the UAVs.

## VI. Acknowledgements

This work was partially supported by the Academy of Finland, 6Genesis Flagship under Grant 318927 and FAITH project under Grant 334280.

## Appendix A
## Jamming Improvement Derivation

Considering $R_S > 0$, the SOP can be expressed as:

$$\begin{aligned} \text{SOP} &= \Pr\left[\log_2\left(\frac{1+\gamma_B}{1+\gamma_E}\right) < R_S\right] \\ &= \Pr\left[g_{AB} < \frac{1}{a}\left(2^{R_S}(1 + b \cdot g_{AE}) - 1\right)\right] \end{aligned} \quad (20)$$

Which is the CDF of $g_{AB}$ averaged over $g_{AE}$, both exponential random variables with parameter $\Omega_{AB}$ and $\Omega_{AE}$ respectively, for which it follows:

$$\begin{aligned} \text{SOP} &= \mathrm{E}_{g_{AE}}\left[F_{g_{AB}}\left(\frac{1}{a}\left(2^{R_S}(1 + b \cdot g_{AE}) - 1\right)\right)\right] \\ &= \mathrm{E}_{g_{AE}}\left[1 - e^{-\Omega_{AB}\frac{1}{a}\left(2^{R_S}(1+b\cdot x)-1\right)}\right] \\ &= 1 - \int_0^\infty e^{-\Omega_{AB}\frac{1}{a}\left(2^{R_S}(1+b\cdot x)-1\right)}\Omega_{AE}e^{-x\Omega_{AE}}dx \end{aligned}$$

After basic derivations, the SOP is given by:

$$\text{SOP} = 1 - e^{-\frac{\Omega_{AB}}{a}\left(2^{R_S}-1\right)}\left(\frac{1}{2^{R_S}\left(\frac{\Omega_{AB}}{\Omega_{AE}}\right)\left(\frac{b}{a}\right) + 1}\right) \quad (21)$$

Then, after some algebraic derivations the jamming improvement $\overline{\Delta}$ expression in (16) follows.